\documentclass[12pt]{amsart}

\long\def\forget#1\forgot{}
\newcommand{\LNF}{\ell_\text{NF}}

\theoremstyle{definition}

\newtheorem{defn}{Definition}
\newtheorem{prob}{Problem}

\newtheorem{algo}{Algorithm}

\newcommand{\inv}{^{-1}}

\begin{document}

\title[Length-based cryptanalysis]{Length-based cryptanalysis: The
case of Thompson's Group}

\begin{abstract}
The length-based approach is a heuristic for solving randomly
generated equations in groups that possess a reasonably behaved
length function. We describe several improvements of the
previously suggested length-based algorithms, which make them
applicable to Thompson's group with significant success rates. In
particular, this shows that the Shpilrain-Ushakov public key
cryptosystem based on Thompson's group is insecure, and suggests
that no practical public key cryptosystem based on the difficulty of
solving an equation in this group can be secure.
\end{abstract}

\author{Dima Ruinskiy}
\author{Adi Shamir}
\author{Boaz Tsaban}
\thanks{The third author is supported by the Koshland Center for Basic Research.}

\address{Faculty of Mathematics,
Weizmann Institute of Science, Rehovot 76100, Israel}
\email{\{dmitriy.ruinskiy, adi.shamir,
boaz.tsaban\}@weizmann.\allowbreak ac.il}

\maketitle

\section{Introduction}

Noncommutative groups are often suggested as a platform for public
key agreement protocols, and much research is dedicated to analyzing
existing proposals and suggesting alternative ones (see, e.g.,
\cite{AAG, Ts1, Ts2, HugTan, Matucci, ShpAssessing, ShpUsh, Unnec},
and references therein).

One possible approach for attacking such systems was outlined by
Hughes and Tannenbaum \cite{HugTan}. This approach relies on the
existence of a good length function on the underlying group, i.e., a
function $\ell(g)$ that tends to grow as the number of generators
multiplied to obtain $g$ grows. Such a length function can be used
to solve, heuristically, \emph{arbitrary} random equations in the
group \cite{Ts1}.

In the case of the braid group, a practical realization of this
approach was suggested in \cite{Ts1}, and the method was extended in
\cite{Ts2} to imply high success rates for subgroups of the
braid group, which are of the type considered in some previously
suggested cryptosystems (e.g., \cite{AAG}).

This \emph{length-based cryptanalysis} usually has smaller success
rates than specialized attacks, but it has the advantage of being
\emph{generic} in the sense that, if there is a good length function
on a group, then the attack applies with nontrivial success rates to
all cryptosystems based on this group (provided that an equation in
the group can be extracted from the public information).

The main problem with existing length-based algorithms is that they
tend to perform well only when the underlying subgroup has few
relations, i.e., it is not too far from the free group. This is not
the case in Richard Thompson's group $F$, since it has a maximal
set of relations: Any nontrivial relation
added to it makes it abelian \cite{CFP96}. In 2004, Shpilrain and
Ushakov proposed a key exchange protocol that uses Thompson's group
$F$ as its platform and reported a complete failure of a
length-based attack on their cryptosystem \cite{ShpUsh}.

In the sequel we introduce several improvements to the length-based
algorithms, which yield a tremendous boost in the success rates for
full size instances of the cryptosystem. The generalized algorithms
presented here are not specific for Thompson's group, and
would be useful in testing the security of any future
cryptosystem based on combinatorial group theoretic problems.

\subsection{History and related works}
The results reported here form the first practical cryptanalysis of
the Shpilrain-Ushakov cryptosystem: The first version of our attack
was announced in the Bochum Workshop \emph{Algebraic Methods in
Cryptography} (November 2005) \cite{RSTBochum}. An improved attack
was announced in the CGC Bulletin in March 2006 \cite{RSTCGC5}.

While we were finalizing our paper for publication, a very elegant
specialized attack on the same cryptosystem was announced by Matucci
\cite{Matucci}. The main contribution of the present paper is thus
the generalization of the length-based algorithms to make them
applicable to a wider class of groups. Moreover, while our general
attack can be easily adapted to other possible cryptosystems based
on Thompson's group, this may not be the case for Matucci's
specialized methods.

\section{The basic length-based attack}

Let $G$ be a finitely generated group with $ S_G =
\{g_1^{\pm1},\dots,g_k^{\pm1}\} $ being its set of generators.
Assume that $x \in G$ is generated as a product, $x = x_1 \cdots
x_n$, where each $x_i \in S_G$ is chosen at random according to some
nontrivial (e.g., uniform) distribution on $S_G$. Assume further
that $w\in G$ is chosen in a way independent of $x$, and that $x,w$
are unknown, but $z=xw\in G$ is known. Suppose that there is a
``length function'' $\ell(g)$ on the elements of $G$, such that with
a nontrivial probability,
\begin{equation*}\label{eq:lenmono}
\ell(x_1\inv z) < \ell(z) < \ell(x_j z)
\end{equation*}
for each $x_j \neq x_1\inv$. To retrieve $x$, we can try to ``peel
off'' the generators that compose it, one by one, using the
following procedure.

\begin{algo}[Length-based attack]\label{algo:lba}
~
\begin{enumerate}
\item Let $j \leftarrow 1$ and $y \leftarrow z$.
\item \label{itm:initstep}For each $g \in S_G$ compute $g\inv y$.
\item\label{tmp} Consider the $h\in S_G$ that minimizes $\ell(h\inv y)$.
(If several such $h$'s exist, choose one arbitrarily or randomly).
\item
\begin{enumerate}
\item If $j=n$, terminate.
\item Otherwise, Let $h_j \leftarrow h$, $j \leftarrow j+1$ and $y \leftarrow
h\inv y$ and return to step \ref{itm:initstep}.
\end{enumerate}
\end{enumerate}
\end{algo}
If $\ell$ is a good length function, then in step \eqref{tmp}, with
some nontrivial probability, $h=x_1$ (or at least $y$ can be
rewritten as a product of $n$ or fewer generators, where $h$ is
the first). It follows that with a nontrivial (though smaller)
probability, $x = h_1h_2\cdots h_n$ after termination.

Instead of assuming that $n$ is known, we can assume that there is a
known, reasonably sized, bound $N$ on $n$, and then terminate the
run after $N$ steps and consider it successful if for some $k\le N$,
$x = h_1\cdot h_2\cdots h_k$. This way, we obtain a short list of
$N$ candidates for $x$. In many practical situations each suggestion
for a solution can be tested, so this is equally good.

In this algorithm, as well as in the ones that follow, the decisions are
\emph{soft} in the sense that if an incorrect generator is chosen
at some stage, this may be repaired later if a generator that cancels it
out (using the group relations) is chosen.

However, in practice the known length functions in many types of
groups are not good enough for
Algorithm \ref{algo:lba} to succeed with noticeable probability.
This is shown in \cite{Ts1}, and is demonstrated further by
the Shpilrain-Ushakov key agreement protocol.

\section{The Shpilrain-Ushakov Key agreement Protocol}

This section is entirely based on \cite{ShpUsh}.

\subsection{Thompson's group}
Thompson's group $F$ is the infinite noncommutative group defined by
the following generators and relations:
\begin{equation}\label{eq:Fpresent}
F = \langle \quad x_0 , x_1 , x_2 , \dots \quad \arrowvert \quad
x_i\inv x_k x_i = x_{k+1} \quad (k>i) \quad \rangle
\end{equation}

Each $w\in F$ admits a unique \emph{normal form}
\cite{CFP96} which has the following structure:
$$w = x_{i_1} \cdots x_{i_r}x_{j_t}\inv \cdots x_{j_1}\inv,$$
where $i_1 \le \cdots \le i_r$, $j_1 \le \cdots \le j_t$, and if
$x_{i}$ and $x_{i}\inv$ both occur in this form, then either
$x_{i+1}$ or $x_{i+1}\inv$ occurs as well. The transformation of an
element of $F$ into its normal form is very efficient: Starting with a
word $w$ of length $n$, the number of required operations is bounded by
a small constant multiple of $n\log n$ \cite{ShpUsh}.

\begin{defn}\label{defn:nflen}
\emph{The normal form length} of an element $w \in F$, $\LNF(w)$, is
the number of generators in its normal form: If the normal form of $w$ is
$x_{i_1} \cdots x_{i_r}x_{j_t}\inv \cdots x_{j_1}\inv$,
then $\LNF(w) = r+t$.
\end{defn}

\subsection{The protocol}\label{subsec:proto}
\begin{enumerate}
\item[(0)] Alice and Bob agree (publicly) on subgroups $A,B,W$ of $F$,
such that $ab=ba$ for each $a \in A$ and each $b \in B$.
\item A public word $w \in W$ is selected.
\item Alice selects privately at random
elements $a_1 \in A$ and $b_1 \in B$, computes $u_1 = a_1 w b_1$, and
sends $u_1$ to Bob.
\item Bob selects privately at random elements $a_2
\in A$ and $b_2 \in B$, computes $u_2 = b_2 w a_2$, and sends $u_2$ to
Alice.
\item Alice computes $K_A = a_1 u_2 b_1 = a_1 b_2 w a_2 b_1$,
whereas Bob computes $K_B = b_2 u_1 a_2 = b_2 a_1 w b_1 a_2$.
\end{enumerate}
As $a_1 b_2 = b_2 a_1$ and $a_2 b_1 = b_1 a_2$, $K_A=K_B$ and so
the parties share the same group element, from which a
secret key can be derived.

\subsection{Settings and parameters}\label{subsec:param}

Fix a natural number $s \ge 2$. Let $S_A = \{x_0 x_1\inv , \dots ,
x_0 x_s\inv\}$, $S_B = \{x_{s+1}, \dots , x_{2s}\}$ and $S_W =
\{x_0, \dots , x_{s+2}\}$. Denote by $A$, $B$, and $W$ the subgroups
of $F$ generated by $S_A$, $S_B$, and $S_W$, respectively. $A$ and
$B$ commute elementwise, as required \cite{ShpUsh}.

Let $L$ be a positive integer. The words $a_1, a_2\in A$, $b_1,
b_2\in B$, and $w\in W$ are all chosen of normal form length $L$, as
follows: Let $X$ be $A$, $B$, or $W$. Start with the empty word, and
multiply it on the right by a (uniformly) randomly selected
generator, inverted with probability $\frac 1 2$, from the set
$S_X$. Continue this procedure until the normal form of the word has
length $L$.

For practical implementation of the protocol, it is suggested in
\cite{ShpUsh} to use $s\in\{3,4,\dots,8\}$ and
$L\in\{256,258,\dots,320\}$.

\section{Success rates for the basic length attack}\label{subsec:attackbasics}

The cryptanalyst is given $w,u_1,u_2$, where $u_1=a_1wb_1$ and $u_2=b_2wa_2$.
This gives rise to $4$ equations:
\begin{eqnarray*}
u_1 & = & a_1wb_1\\
u_2 & = & b_2wa_2\\
u_1\inv & = & b_1\inv w\inv a_1\inv\\
u_2\inv & = & a_2\inv w\inv b_2\inv
\end{eqnarray*}
He can apply Algorithm \ref{algo:lba} to each equation, hoping
that its leftmost unknown element will appear in the resulting
list of candidates. Note that even a single success out of the $4$
runs suffices to find the shared key.

Here $n$, the number of generators multiplied to obtain each
element, is not known. We took the bound $2L$ on $n$, as experiments
show that the success probability does not increase noticeably when
we increase the bound further. This is the case in all experiments
described in this paper.

Experiments show that the success probability of finding $a_1$ given
$a_1wb_1$ is the same as that of finding $a_2\inv$ given $a_2\inv
w\inv b_2\inv$, that is, the usage of the same $w$ in both cases does not
introduce noticeable correlations. A similar assertion
holds for $b_2$ and $b_1\inv$. We may therefore describe the task in
a compact manner:
\begin{quote}
Given $awb$, try to recover either $a$ or $b$.
\end{quote}
The probabilities $p_a,p_b$ of successfully recovering $a$ and $b$
(respectively) induce the total success rate by
$1-(1-p_a)^2(1-p_b)^2$.

The attack was tested for the minimal recommended value $s=3$, and
for the cut-down lengths $L\in\{4,8,\dots,128\}$. (Each attack in
this paper was tested against at least 1000 random keys, in order to
evaluate its success rates.)

The results, presented in Table \ref{tbl:basic},
show that this is not a viable attack: The recommended
parameter is $L\ge 256$, and already for $L=128$ the attack failed in all of our tries.

\begin{table}[!htp]
\caption{Success rates for the basic length attack ($s=3$)}\label{tbl:basic}
\begin{tabular}{|r|r|r|r|}
\hline
$L$ & $a$ recovery & $b$ recovery & Total \\
\hline
4 &  88.4\% &  82.6\% &  99.96\% \\
8 &  62.3\% &  56.2\% &  97.3\% \\
16 &  29.1\% &  26.9\% &  73.1\% \\
32 &  10.2\% &  8.2\% &  32\% \\
64 &  0.9\% &  1\% &  3.7\% \\
128 & 0\% & 0\% & 0\% \\
\hline
\end{tabular}
\end{table}

\section{Using memory}

To improve the success rates, it was suggested in \cite{Ts2} to
keep in memory, after each step, not only the element that yielded the
shortest length, but a fixed number $M>1$ of elements with the
shortest lengths among all tested elements. Then, in the next step, all possible extensions of
each one of the $M$ elements in memory with each one of the generators are
tested and again the best $M$ elements among them are kept (see \cite{Ts2} for
a formal description of this algorithm).

The time and space complexities of this attack increase linearly
with $M$. The previous length-based attack is the special case of
the memory attack, where $M=1$. Except for pathological cases, the
success rates increase when $M$ is increased. See \cite{Ts2} for
more details.

We have implemented this attack against the minimal recommended
parameters $s=3, L=256$, and with each $M\in\{4,16,64,256,1024\}$.
The success rates appear in Table \ref{tbl:basicmemory}.

\begin{table}[!htp]
\caption{Success rates for the basic length attack with
memory ($s=3, L=256$)}\label{tbl:basicmemory}
\begin{tabular}{|r|r|r|r|}
\hline
$M$ & $a$ recovery & $b$ recovery & Total \\
\hline
$\leq 64$ &  0\% &  0\% &  0\% \\
256 & 1.5\% & 0.1\% & 3.2\% \\
1024 & 5.7\% & 0.1\% & 11.3\% \\
\hline
\end{tabular}
\end{table}

We see that $M$ must be rather large in order to obtain high success rates.
The experiments in \cite{Ts2} yielded much higher success
rates for braid groups. The reason for this seems to be that the
length-based approach is more suitable for groups which have few
relations (i.e., are close to being free) \cite{Ts1}, whereas here
the underlying groups have many relations. The next section shows
how to partially overcome this problem.

\section{Avoiding repetitions}

During the run of the algorithm described in the previous section,
we keep a hash list. Before checking the length score of an element,
we check if it is already in the hash list (i.e., it has been
considered in the past). If it is, we drop it from the list of candidates.
Otherwise, we add it to the hash list and proceed as usual.

In the case $M=1$, this forces the algorithm not to get into loops.
Thus, this improvement can be viewed as a generalization of avoiding
loops to the case of arbitrary $M$.

\subsection{Results}\label{MattackOriginal}
The results for $s=3,L=256$ are summarized in Table \ref{tbl:matk2}.

\begin{table}[!htb]
\caption{Success rates for repetition-free memory attack ($s=3,L=256$)}\label{tbl:matk2}
\begin{tabular}{|r|r|r|r|r|}
\hline
$M$ & $a$ recovery & $b$ recovery & Total\\
\hline
4 & 0\% & 0\% & 0\% \\
16 & 2.3\% & 1.1\% & 6.6\% \\
64 & 10.8\% & 2.3\% & 24\%\\
256 & 14.3\% & 3.8\% & 32\%\\
1024 & 20.4\% & 11\% & 49.8\% \\
\hline
\end{tabular}
\end{table}

It follows that our improvement
is crucial for the current system:
Compare $50\%$ for $M=1024$ in
Table \ref{tbl:matk2} to the $11\%$ for the same
$M$ obtained in Table \ref{tbl:basicmemory}
before we have discarded repetitions.

A success rate of $50\%$ should be considered a complete
cryptanalysis of the suggested cryptosystem. We will, however,
describe additional improvements, for two reasons.

\subsubsection*{Generality}
The Shpilrain-Ushakov cryptosystem is just a test
case for our algorithms. Our main aim is to obtain generic algorithms
that will also work when other groups are used, or when Thompson's
group is used in a different way.

\subsubsection*{Iterability}
As pointed out by Shpilrain \cite{ShpAssessing}, there is a very
simple fix for key agreement protocols that are broken with
probability less than $p$: Agree on $k$ independent keys in
parallel, and XOR them all to obtain the final shared key. The probability
of breaking the shared key is at most $p^k$. In other words, if a
system broken with probability $p_0$ or higher is considered
insecure, and $k$ parallel keys are XORed, then the attack on a
single key should succeed in probability at least ${p_0^{1/k}}$. If
we consider a parallel agreement on up to $100$ keys practical, and
require the probability of breaking all of them to be below
$2^{-64}$, then we must aim at a success rate of at least
$2^{-64/100}\approx 64\%.$ For $p_0=2^{-32}$, we should aim at
$80\%$.

\section{Interlude: Memory is better than look-ahead}

An alternative extension of the basic attack is obtained by testing
in each step not just the $2k$ generators in $S_G$, but all the
$(2k)^t$ $t$-tuples of generators $g_{i_1}^{\pm 1} \cdots
g_{i_t}^{\pm1}$. After computing the length of each of the peeled-off results,
one takes only the first generator of the leading $t$-tuple,
and repeats the process. This is called \emph{look-ahead
of depth $t$} \cite{HugTan, Ts1}. The complexity of this approach
grows exponentially with $t$.

In order to compare this approach with the memory approach, we
should compare attacks using roughly the same number of operations.
The products of all possible $t$-tuples can be precomputed, so that
each step requires $(2k)^t$ group multiplications. In the memory
attack, each step requires $M\cdot 2k$ group multiplications. Thus,
look-ahead of depth $t$ should be compared to $M=(2k)^{t-1}$.

\subsection{Results}
The look-ahead attack was tested for $s=3$, $L=256$. We tried $t\in\{2,3,4\}$,
which correspond to $M\in\{6,6^2,6^3\}$, respectively. The results
are presented in Table \ref{tbl:lahead}. For $t=3,4$, we have also
tried the intermediate approach where a look-ahead of depth $t-i$ is
performed ($i=1,2$) for each member of the list and $M=(2k)^i$.

\begin{table}[!htp]
\caption{Success rates for look-ahead LA, memory attack M, and combined M\&LA ($s=3,L=256$)}\label{tbl:lahead}
\begin{tabular}{|rrr||r|r||r|r||r|r|r|}
\hline & & & \multicolumn{2}{c||}{$a$ recovery} &
\multicolumn{2}{c||}{$b$ recovery} & \multicolumn{3}{c|}{Total}\\
\hline
$t$ & $M$ & $t,M$ & LA & M & LA & M & LA & M & M\&LA\\
\hline
2 & 6 & ~---~ & 0\% & 0.1\% & 0\% & 0.6\% & 0\% & 1.4\% & ~---~ \\
3 & 36 & 2,6 & 0.1\% & 7.4\% & 0.1\% & 3.6\% & 0.4\% & 20.3\% & 6.8\% \\
4 & 216 & 2,36 & 1.4\% & 16.8\% & 0.8\% & 8.3\% & 4.3\% & 41.8\% & 31.2\% \\
   & &    3,6  &       &        &       &       &        &         & 14.4\%\\
\hline
\end{tabular}
\end{table}

It follows that increasing $M$ is always better than using look-ahead of similar complexity.
This was also observed in \cite{Ts1, Ts2} for other settings.

\section{Automorphism attacks}

Recall our problem briefly: $G=\langle S_G\rangle$, where $S_G =
\{g_1^{\pm1},\dots,g_k^{\pm1}\}$. $x,w \in G$ are unknown and chosen
independently, and $z=xw\in G$ is known. We wish to find (a short
list containing) $x$. Write $x=h_1\cdots h_n$.

Let $\varphi$ be an automorphism of $G$. Applying $\varphi$, we have
that $\varphi(z)=\varphi(x)\varphi(w)$, and
$\varphi(x)=\varphi(h_1)\cdots \varphi(h_n)$. This translates the
problem into the same group generated differently: $G=\langle
\varphi(S_G)\rangle$, where $\varphi(S_G) =
\{\varphi(g_1)^{\pm1},\dots,\varphi(g_k)^{\pm1}\}$. Solving the
problem in this group to find $\varphi(x)$, gives us $x$.

Solving the problem in the representation of $G$ according to $\varphi$
is equivalent to solving the original problem with the alternative
length function
$$\ell_\varphi(w)=\ell(\varphi(w)).$$
Indeed,
$$\ell(\varphi(g_i)^{\pm1}\varphi(x)\varphi(w))=
\ell(\varphi(g_i^{\pm1}xw))=\ell_\varphi(g_i^{\pm1}xw).$$

It could happen that a certain key which is not cracked by a given
length attack using a length function $\ell$, would be cracked
using $\ell_\varphi$.

If we choose $\varphi$ at ``random'' (the canonical example being an
inner automorphism $\varphi(w)=g\inv w g$ for some ``random'' $g$),
we should expect smaller success rates, but on the other hand the
introduced randomness may be useful in one of the following ways.
Let $\Phi$ be a finite set of automorphisms of $G$.

\subsubsection*{Average length attack.}
We can take the \emph{average} length
$$\ell_\Phi(w)=\frac{1}{|\Phi|}\sum_{\varphi\in\Phi}\ell_\varphi(w).$$
If the elements $\varphi$ of $\Phi$ are chosen independently
according to some distribution, then
$$\lim_{|\Phi|\to\infty}\ell_\Phi(w)=E(\ell_\varphi(w)),$$
where the expectancy is with regards to the distribution of the chosen
elements $\varphi$.
This approach should be useful when the length function
$\ell_E(w)=E(\ell_\varphi(w))$ is good.
This would be the case if there are only weak correlations between the
different length functions:
Roughly speaking, if there are weak correlations
between the different length functions $\ell_\varphi$, and for a
random $\varphi$ the probability of getting a correct generator is
some $p$ with $\epsilon=p-(1-p)>0$, then for
$|\Phi|=O(1/\epsilon^2)$, a correct generator will get the the
shortest average length $\ell_\Phi$ almost certainly.

\subsubsection*{Multiple attacks}\label{ma}
Write $\Phi=\{\varphi_1,\dots,\varphi_m\}$. We can attack the key using
$\ell_{\varphi_1}$. If we fail, we attack the same key again using
$\ell_{\varphi_2}$, etc. Here too,
if there are weak correlations between the different length functions
and $|\Phi|$ is large, then we are likely to succeed.

In the case of Thompson's group $F$, the family of automorphisms
is well understood (they are all conjugations by elements of
some well defined larger group) \cite{Brin96}. However, since
we are interested in ``generic'' attacks, we considered
only inner automorphisms.

\subsection{Results}\label{manyaut}
All experiments were run for parameters $s=3,L=256$ and without
memory extensions ($M=1$). All conjugators defining the inner automorphisms
were random elements of length $64$. The complexity of the two
described attacks is similar to that of the memory attack with
$M=|\Phi|$.

\subsubsection*{Average length attack.}
We tried the average length attack with $|\Phi| \in
\{4,16,64,256,1024\}$. Not a single one of the experiments was
successful. This implies either that the correlation between the
different length functions is rather high or that the actual success
probability for a given length function is very low.

\subsubsection*{Multiple attacks}
The success rates appear in Table \ref{tbl:multiple}.

\begin{table}[!htp]
\caption{Success rates for the multiple attack ($s=3,L=256$)}\label{tbl:multiple}
\begin{tabular}{|r|r|r|r|}
\hline
$|\Phi|$ & $a$ recovery & $b$ recovery & Total \\
\hline
4 &  0.1\% &  0\% &  0.2\% \\
16 &  0.9\% &  0\% &  1.8\% \\
64 &  2.2\% &  0\% &  4.4\% \\
256 & 2.2\% & 0\% & 4.4\% \\
1024 & 2.5\% & 0\% & 4.9\% \\
\hline
\end{tabular}
\end{table}

While an improvement is observed, it is also seen that there remain
substantial correlations and the success rate does not increase fast
enough when $|\Phi|$ is increased.
Comparing the results to those in Table \ref{tbl:matk2}, we
see that in the current setting, increasing the memory is far better
than using many automorphisms.

\section{Alternative solutions}\label{sec:alt}

Thus far, we have concentrated on the problem: Given $w$ and $awb$,
find the \emph{original} $a$, or rather, a short list containing
$a$. But as Shpilrain and Ushakov point out \cite{Unnec}, it
suffices to solve the following problem.

\begin{prob}[Decomposition]\label{prob:decomp}
Given $w\in F$ and $u=awb$ where $a\in A$ and $b\in B$, find some
elements $\tilde a\in A$ and $\tilde b\in B$, such that $\tilde
aw\tilde b=awb$.
\end{prob}

Indeed, assume that the attacker, given $u_1 = a_1 w b_1$, finds
$\tilde a_1 \in A$ and $\tilde b_1 \in B$, such that $\tilde a_1
w \tilde b_1 = a_1 w b_1$. Then, because $u_2 = b_2 w a_2$ is known,
the attacker can compute
$$\tilde a_1 u_2 \tilde b_1 = \tilde a_1 b_2 w a_2 \tilde b_1 = b_2
\tilde a_1 w \tilde b_1 a_2 = b_2 u_1 a_2 = K_B,$$
and similarly for $b_2wa_2$.

Consider Problem \ref{prob:decomp}. To each $\tilde a\in A$ we can
compute its \emph{complement} $\tilde b = w\inv \tilde a\inv u =
w\inv \tilde a\inv(awb)$, such that $\tilde a w\tilde b = awb$. The
pair $\tilde a,\tilde b$ is a solution to this problem if, and only
if, $\tilde b\in B$. A similar comment applies if we start with
$\tilde b\in B$. This involves being able to determine whether
$\tilde b\in B$ (or $\tilde a\in A$ in the second case). This
\emph{membership decision problem} turns out to be trivial in our case.

$A$ is exactly the set of all elements in $F$, whose normal form
is of the type
\begin{equation*}
x_{i_1} \dots x_{i_m} x_{j_m}\inv \dots x_{j_1}\inv,
\end{equation*}
i.e., positive and negative parts are of the same length, and in
addition $i_k-k<s$ and $j_k-k<s$ for every $k=1,\dots,m$. $B$
consists of the elements in $F$, whose normal form does not contain
any of the generators $x_0,x_1,\dots,x_s$ (or their inverses)
\cite{ShpUsh}. In both cases, the conditions are straightforward to
check.

Following is an algorithm for solving Problem \ref{prob:decomp},
which incorporates the new flexibility into the halting rule.

\begin{algo}[Alternative solution search]\label{algo:equiv}
~
\begin{enumerate}
\item Execute Algorithm \ref{algo:lba} (with any of the introduced
extensions), attempting to recover $a$.
\item For each candidate (prefix) $\tilde a$ encountered during any
step of the algorithm, compute the complement $\tilde b = w\inv
\tilde a\inv u$.
\item\label{success} If $\tilde b \in B$, halt.
\end{enumerate}
\end{algo}

Note that if the algorithm halts in step \eqref{success}, then
$\tilde a,\tilde b$ is a solution for the decomposition problem.

The above procedure can be executed separately for each of the four
given equations. It suffices to recover a single matching pair
in any of the four runs to effectively break the cryptosystem.

\subsection{When the group membership problem is hard}
It should be stressed that solving the group membership is not
necessary in order to cryptanalyze the system. Indeed, given
$u_1=a_1wb_1$ and $u_2=b_2wa_2$, we can apply Algorithm
\ref{algo:equiv} to, e.g., $u_1=a_1wb_1$, replacing its step
\eqref{success} by checking whether the suggested key $\tilde a u_2
\tilde b$ succeeds in decrypting the information encrypted between
Alice and Bob. Our experiments showed that for all reasonable
parameters, this formally stronger attack has the same success
rates. However, this alternative approach is useful in
other groups, in which the membership problem is difficult.

\subsection{Results}
We have repeated all major experiments for $s=3,L=256$,
but this time considered each alternative solution
a success. We consider only the repetition-free versions of the
attacks, as they are much more successful.

\subsubsection*{Average automorphism attack}
While being substantially better than the $0\%$ reported
in Section \ref{manyaut} before allowing alternative solutions,
the results here are still not satisfactory: For all $|\Phi|\in\{4,16,\dots,1024\}$,
the average rates were close to $17\%$. This suggests that in this setting,
the average length converges to the expected length very quickly.

\subsubsection*{Multiple attack}
The success rates for the multiple attack (page \pageref{ma})
are quite good when alternative solutions are accepted, as shown in Table \ref{tbl:multiple2}.
\begin{table}[!htp]
\caption{Success rates for the multiple attack ($s=3,L=256$)}\label{tbl:multiple2}
\begin{tabular}{|r|r|r|r|}
\hline
$|\Phi|$ & $a$ recovery & $b$ recovery & Total \\
\hline
4 &  7.1\% & 13.7\% & 35.7\% \\
16 &  11.3\% & 20.4\% & 50.1\% \\
64 &  11.5\% & 23.3\% & 53.9\% \\
256 & 16.7\% & 24.5\% & 60.4\% \\
1024 & 14.5\% & 20.2\% & 53.4\% \\
\hline
\end{tabular}
\end{table}

It is observed, though, that no significant
improvement is obtained when moving from $|\Phi|=256$ to $|\Phi|=1024$
(what looks in the table like a drop in the probability is probably a statistical fluctuation,
but it still shows that the real probability does not increase substantially).

\subsubsection*{Memory attack}
This attack, which corresponds to Section \ref{MattackOriginal}
but allows alternative solutions, gives the best results on the studied case.
We have tried it against the minimal suggested parameters
($s=3,L=256$), as well as the maximal suggested parameters
($s=8,L=320$). The results appear in Table \ref{MattackAlter}.

\begin{table}[!htp]
\caption{Success rates for memory attack with alternative solutions}\label{MattackAlter}

\begin{tabular}{|r||r|r|r||r|r|r|}
\hline & \multicolumn{3}{c||}{$s=3,L=256$} &
\multicolumn{3}{c|}{$s=8,L=320$} \\
\hline
$M$ & $a$ & $b$ & Total & $a$ & $b$ & Total \\
\hline
1    &  9.3\% &  5.3\% & 26.2\% &  8.0\% &  6.1\% & 25.4\% \\
4    & 12.1\% &  7.4\% & 33.7\% & 10.9\% & 10.9\% & 37.0\% \\
16   & 15.6\% & 10.9\% & 43.4\% & 11.3\% & 11.5\% & 38.4\% \\
64   & 27.8\% & 14.7\% & 62.1\% & 17.3\% & 13.1\% & 48.4\% \\
256  & 35.8\% & 20.1\% & 73.7\% & 18.0\% & 15.3\% & 51.8\% \\
1024 & 41.5\% & 25.0\% & 80.7\% & 22.2\% & 14.5\% & 55.8\% \\
\hline
\end{tabular}
\end{table}

Note that for $s=3,L=256$, we have that $M=16$ with alternative solution search
gives success rates almost equal to those of $M=1024$ (which is $64$ times slower)
without it, and that $M=1024$ with alternative solution search results in success
rate of about $80\%$.

It is also interesting to observe that while increasing the parameters
reduces the success rates, the success rates are significant
even when the maximal recommended parameters are taken.

Based on Table \ref{MattackAlter}, we conclude that
the Shpilrain-Ushakov cryptosystem is broken, even if
iterated up to one hundred times.

\section{Conclusions}

We have described several improvements on the standard
length based attack and its memory extensions. They include:
\begin{enumerate}
\item \label{itm:repetitions} Avoiding repetitions, which is
especially important in groups such as Thompson's group $F$,
that are far from being free;
\item \label{itm:automorphism} Attacking each key multiple times, by
applying each time a random automorphism, or equivalently taking the
length function induced by such automorphisms;
\item \label{itm:alternative} Looking for alternative solutions
which are not necessarily the ones used to generate the equations.
\end{enumerate}
We have tested these improvements against the Shpilrain-Ushakov
cryptosystem, and in this case each of them increased the success
probability substantially, with \eqref{itm:repetitions} being
somewhat better than \eqref{itm:automorphism}, and
\eqref{itm:alternative} being a useful addition to any of these.
It could be that for other cryptosystems,
\eqref{itm:automorphism} will prove to be better than \eqref{itm:repetitions}.

The important advantage of our approach is that it is generic and
can be easily adjusted to any cryptosystem based on a group that
admits a reasonable length function on its elements. As such, we
believe that no cryptosystem leading to
equations in a noncommutative group can be considered secure before
tested against these attacks.

It is a fascinating challenge to find an alternative platform
group where the attacks presented here fail.
Such a platform may exist, and the methods presented
here should be useful for dismissing many of the insecure
candidates.

\subsection*{Acknowledgements}
We thank Francesco Matucci for his useful comments on this paper.

\end{document}